\documentstyle[epsfig]{lamuphys}
\makeatletter
\let\chapter\hid@chapter
\makeatother
\begin{document}
\pagenumbering{arabic}
\title{Peculiar Motions Of Clusters In The Perseus--Pisces Region} %

\author{R.J.\,Smith\inst{1},  M.J.\,Hudson\inst{1,2},
J.R.\,Lucey\inst{1} and
J.\,Steel\inst{1}}

\institute{Department of Physics, University of Durham,
South Road, Durham DH1 3LE, U.K.
\and
Department of Physics and Astronomy, University of Victoria, P.O. Box 3055,
Victoria BC V8W 3PN, Canada}

\maketitle

\begin{abstract}
We present results of a new study of peculiar motions of 7 clusters
in the Perseus--Pisces (PP) region, using the Fundamental Plane as a distance indicator. The sample
is calibrated by reference to 9 additional clusters with data from the literature.
Careful attention is paid to the matching of spectroscopic and photometric data from several 
sources.

For six clusters in the PP supercluster no significant peculiar motions are detected.
For these clusters we derive a bulk motion of 60$\pm$220 km\,s$^{-1}$, in the CMB frame, 
directed towards the Local Group. This non-detection is in marginal conflict with previous 
Tully-Fisher studies. Two clusters in the background of the supercluster exhibit 
significant negative peculiar velocities, characteristic of backside infall into PP.

A bulk-flow fit to all 16 clusters reveals a statistically insignificant 
motion of 430$\pm$280 km\,s$^{-1}$ 
towards $l=265^{\circ}, b=26^{\circ}$ (CMB frame). Comparison with the velocity field
predicted from the IRAS 1.2Jy survey yields $\beta = 1.0 \pm 0.5$. We find no evidence for residual bulk motions generated by mass concentrations
beyond the limiting depth of the IRAS density field.

\end{abstract}
\section{Introduction}

The Perseus--Pisces (PP) supercluster, at $cz \sim$5000\,km\,s$^{-1}$, lies
directly opposite the apex of the large-scale streaming detected by \cite{lb88}.
If, as in the flow model of \cite{fb88}, the local dynamics are dominated
by a single `Great Attractor' (GA) beyond Hydra--Centaurus, then the peculiar
velocity at PP is predicted to be $\la 150$\,km\,s$^{-1}$. If however, 
very distant 
sources are responsible for large-scale motions, then the PP region
would be expected
to share in a bulk streaming velocity of $\sim 500$\,km\,s$^{-1}$.

Measurements of bulk motion in PP have been made by \cite{w91} and 
by \cite{ha:mo}. These studies, employing the Tully--Fisher (TF) relation
for spiral galaxies, both argue for a large bulk motion
($\sim 400$\,km\,s$^{-1}$). \cite{cf93} support this result, and invoke
very large-scale density fluctuations to account for the large coherence 
length of the flow.

Motions in the PP region have not been well-studied using elliptical galaxy 
distance  indicators. Here, we describe a programme to measure 
peculiar motions for 7 clusters in PP, using the Fundamental Plane 
(FP) relation for early-type galaxies. Spectroscopic
and photometric data and methods are presented in full by \cite{pp-i}. 
\cite{pp-ii} present a detailed account of the FP fits and 
velocity field analyses.

\section{A new survey} \label{methodology}

The current survey is based on a sample of early-type galaxies in clusters.
Clusters are chosen since {\it a priori} grouping in redshift-space reduces
the magnitude of Malmquist bias effects. This is especially desirable in regions
of high density constrast, such as PP, where corrections for inhomogeneous Malmquist
bias would otherwise be large and uncertain.
Early-type galaxies provide the best means of 
probing cluster cores with minimal contamination from field galaxies.

\begin{figure} 
\epsfig{file=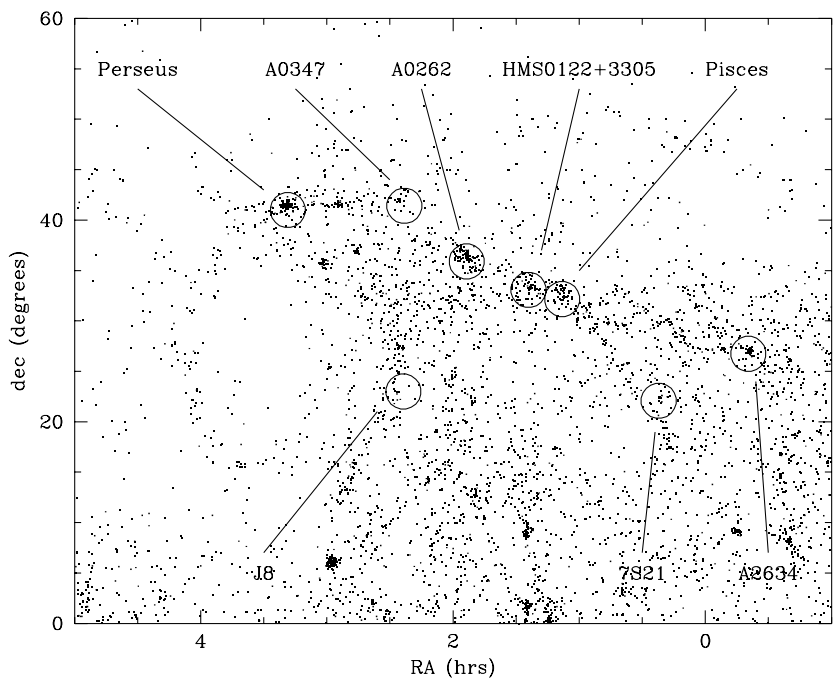,height=120mm,width=120mm}
\caption{Projected distribution of $cz < 12000$\,km\,s$^{-1}$ galaxies in the
PP region, selected from the ZCAT compilation of Huchra et al. (1993). 
Open circles mark the location of clusters observed in this programme.}  \label{skyproj}
\end{figure}

Figure~\ref{skyproj} shows the projected distribution of the cluster sample,
along with the projected galaxy distribution.
The supercluster `ridge' at $cz \sim 5000$\,km\,s$^{-1}$ is evident in the plot, and is
traced by six observed clusters (Perseus, A0347, A0262, HMS0122+3305, Pisces and 7S21).
The sample includes the cluster J8, located in the background of the ridge, 
at $cz \sim 9000$\,km\,s$^{-1}$. The cluster A2634 is formally part of our calibration
sample, but can also be considered as a PP background cluster.

\subsection{Photometry}

The fundamental plane parameters $R_{\rm e}$ and $\langle\mu\rangle_{\rm e}$
are measured from R-band photometric observations obtained at
the 1m Jacobus Kapteyn 
Telescope on La Palma. 
Standard reduction techniques are applied, with parameters finally being 
derived from an $R^{1/4}$-law fit to the aperture magnitude growth-curve. 
Internal and external comparisons indicate that measurement uncertainties
in the 
photometric parameters contribute distance errors smaller than 1.5\%
per galaxy. 

\subsection{Spectroscopy} 

Velocity dispersions, $\sigma$, are derived from spectra obtained at
the 2.5m Isaac Newton Telescope on La Palma. Internal comparisons
indicate an average uncertainty of 7.6\% per measurement. The 
aperture correction of \cite{jfk2} is adopted, with all
$\sigma$ measurements referred to a physical aperture of diameter
1.19$\,h^{-1}$\,kpc.

In order to construct large samples of peculiar velocity data, it 
is necessary to merge measurements from different observing runs,
instrumental configurations, observers, etc.
Despite careful attempts to correct for aperture effects, systematic
offsets often persist at the 3-4\% level between velocity dispersions
measured on different systems. Whilst the precise cause of these offsets
is not fully understood, it is vital that they are corrected for and (most
importantly) that the uncertainties on these corrections are included in
the overall error budget.

The removal of systematic offsets can be achieved by intercomparison
of measurements for galaxies common to two or more systems. 
The set of overlap observations used here includes 1300
measurements for 350 different galaxies, on 19 systems. 
We adopt the (aperture-corrected) Lick system of \cite{dav87} as our 
`standard system'. For all other systems,
we determine corrections required to match onto this standard. Since
many galaxies have data on more than two systems, the fit has been 
performed simultaneously for all the corrections. 

From bootstrap realisations of this process, we estimate the error
on each system correction and also the correlation between the
errors for different systems. The mean systematic error in $\sigma$ can
then be computed for each cluster. For the PP ridge 
clusters, this `system matching' error is $\sim 100$\,km\,s$^{-1}$. 

\subsection{Fundamental Plane fits}

Since the PP cluster sample studied here has very limited sky coverage, 
it would be impossible, given the present data alone, to differentiate 
between a bulk motion of PP and a velocity zero-point error. In order
to resolve this degeneracy, we utilise data for a well-distributed sample
of `calibration clusters'. Specifically, we include six rich clusters
(A0194, A0539, A3381, A3574, DC2345-28, Hydra) from the work of 
\cite{jfk3}, and
three well-studied clusters (Coma, A2199, A2634) from \cite{lu:gu}.
Photometric parameters from the three sources can be brought onto a
consistent system by the application of well defined offsets (Smith
et al. 1997a).

Using the resulting sample of 16 clusters, we fit
simultaneously for the global FP slope parameters,
and for the individual cluster zero-points. The fit is performed by 
minimizing residuals in $\log\sigma$. This `inverse fit' is unbiased by
selection on photometric parameters (radius, surface brightness
or any combination thereof, such as total magnitude).
The inverse fit would, however, be biased by any explicit selection on
$\log \sigma$. In the present work, no such selection occurs,
since galaxies are not thrown out of the sample, {\it
a posteriori}, based on their measured velocity dispersions.

The best fitting fundamental plane for the sample is given by 
\[
\log R_{\rm e} = 1.383 ( \pm 0.040 )\log\sigma + 0.326 (\pm 0.011)\langle\mu\rangle_{\rm e} + \gamma_{\rm cl}
\]
with scatter equivalent to a distance error of 20\% per galaxy. 
The cluster zero-points, $\gamma_{\rm cl}$, are used to derive relative 
distances, as dicussed below. In practice we find that an initial 
calibration, based on the assumption of zero peculiar motion for 
Coma, leads to a mean radial peculiar velocity for the sample
which is indistinguishable from zero. Hereafter, we adopt this
simply-defined zero-point for the present analysis.

\begin{figure} 
\epsfig{file=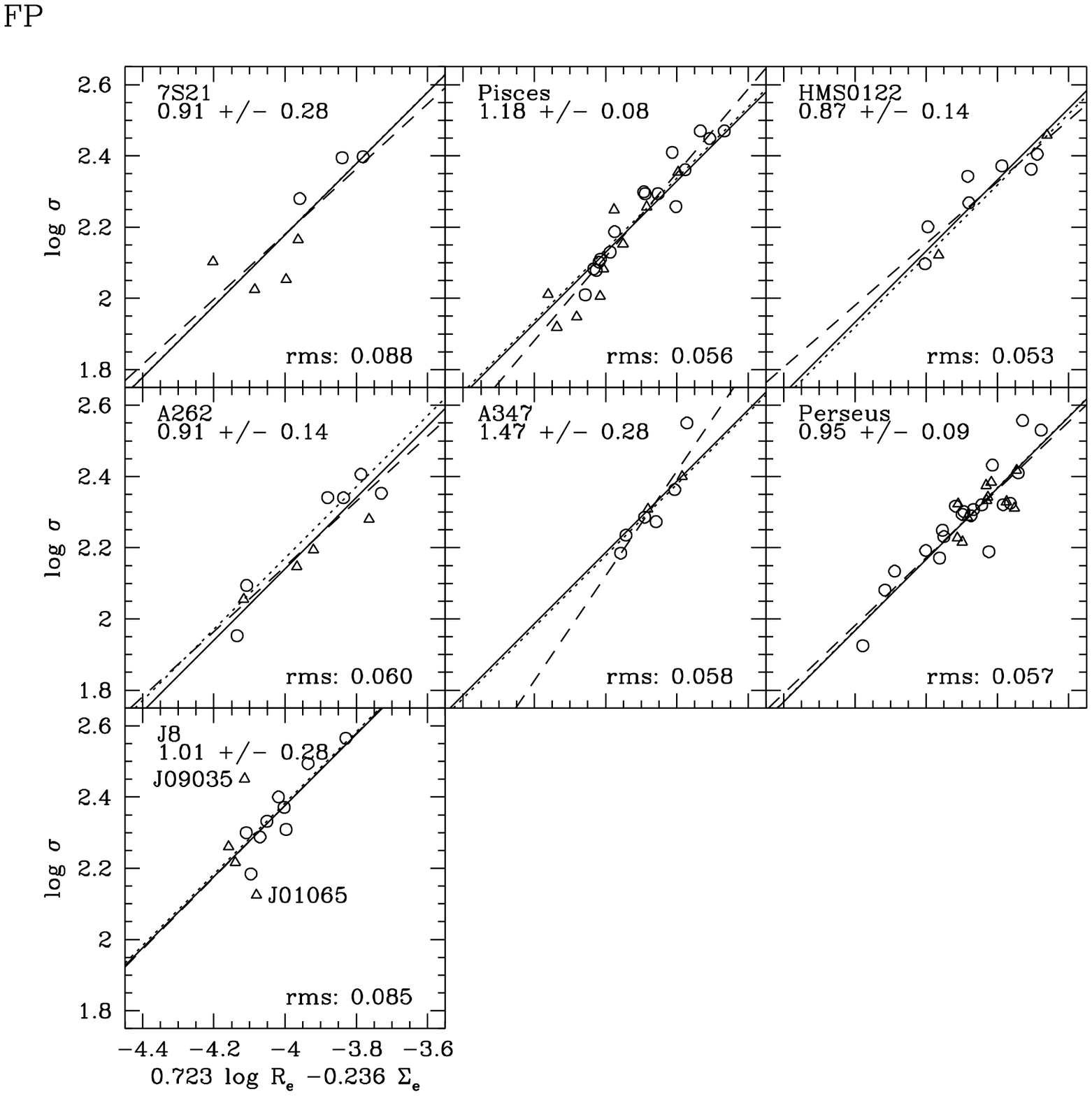,height=120mm,width=120mm}
\caption{
  FP data and fits for the clusters 7S21, Pisces, HMS0122, A0262, A0347,
  Perseus and J8.  Early type galaxies (E, E/S0, D or
  cD) are indicated by circles, later types by
  triangles.  The solid line shows the global inverse FP, found by
  minimising $\sigma$ residuals simultaneously over the whole cluster
  sample with the same slope but varying zero-points for each cluster.
  Each cluster's measured scatter in $\sigma$ around this global fit
  is given in the lower right-hand corner.  Galaxies which deviate from
  the global fit by more than 2.5 times the global scatter are
  labelled.  The dotted line shows the median of the residuals with
  the slope fixed from the whole cluster sample.  The dashed line
  shows the best slope and zero-point fit to the individual cluster.
  The individual cluster slope relative to the global slope is given
  in the upper left-hand corner.}
\label{fpplots1}
\end{figure}

\begin{figure} 
\epsfig{file=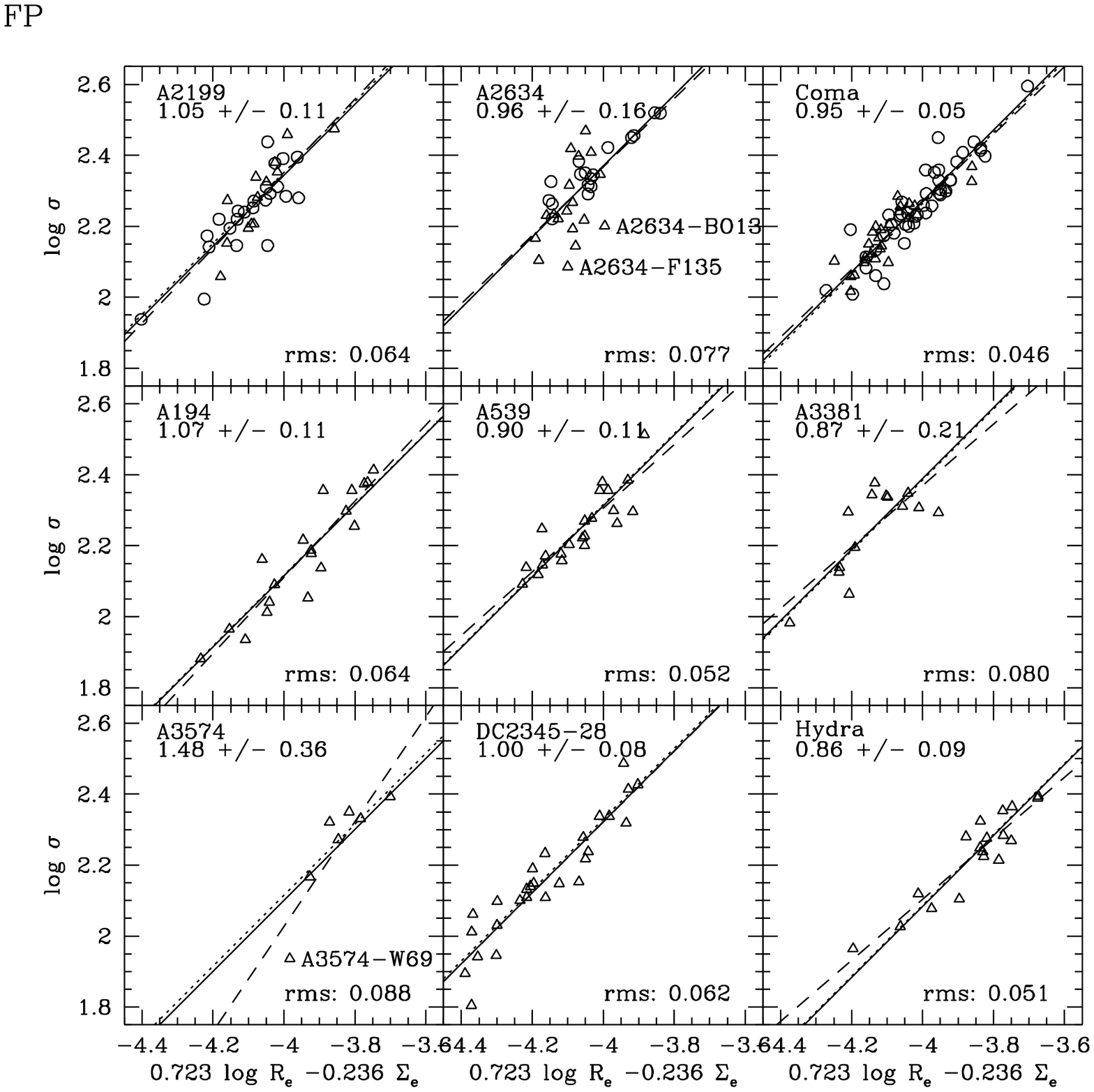,height=120mm,width=120mm}
\caption{
  FP data and fits for the calibration clusters A2199, A2634,
  Coma, A0194, A0539, A3381, A3574,
  DC2345-28 and Hydra. Symbols are as in   Figure 2 for
Coma, A2199 and A2634. For other clusters, all galaxies are indicated by 
triangles, irrespective of type. Lines are as in Figure 2}
\label{fpplots2}
\end{figure}

In Figures~\ref{fpplots1} and ~\ref{fpplots2} we show the FP fits for the 16 clusters 
separately. The slopes derived for individual clusters are consistent
in every case with the slope of the globally-defined FP.

\section{Results} \label{results}

\subsection{Peculiar velocities}

\begin{figure} 
\epsfig{file=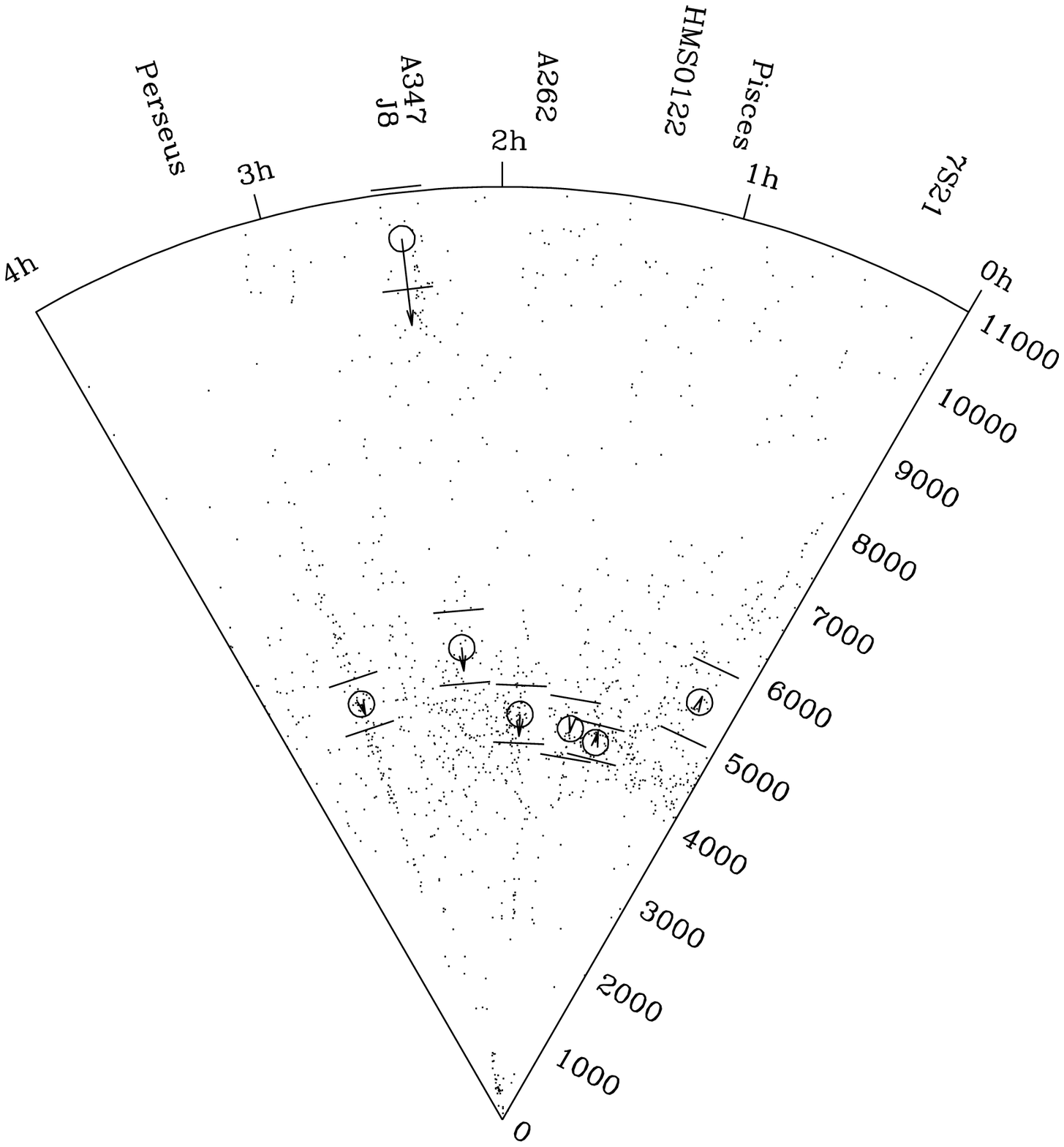,height=115mm,width=115mm}
\caption{Cone diagram showing peculiar velocity vectors deduced from this study.
Small points indicate the CMB-frame redshift-space positions of galaxies from
the ZCAT compilation (Huchra et al. 1993). PP clusters are shown at their inferred distances by 
open circles, with vectors extending to their CMB-frame velocities to indicate 
peculiar motions. $1\sigma$ position errors are indicated by the bracketing 
lines. The six ridge clusters have peculiar motions indistinguishable from
zero. In the background of the supercluster, J8 has a significant 
peculiar velocity, as does A2634 which lies just outside the range of this plot.} \label{cone1}
\end{figure}

Distances and peculiar velocities for the 16 cluster sample are 
presented in Table~\ref{pecvels}. Here, the distance estimates have been 
corrected for homogeneous Malmquist bias. The use of a cluster sample leads
to corrections of only 0.5--2\% for this effect. 
Inhomogeneous Malmquist bias corrections are expected to be smaller than this,
and are neglected here. The velocity vectors for clusters 
in PP are shown in graphical form in Figure~\ref{cone1}.

The present data do not indicate a significant peculiar motion
for any of the six clusters in the ridge of PP. In the background of the 
supercluster, J8 and A2634 display large, significant peculiar velocities
in the sense expected if they are involved in `backside infall' into PP.

In Table~\ref{velcomps}, we present comparisons between the current results
and peculiar velocities measured by other authors. The agreement is good for 
most clusters. 

\begin{table*}
\caption{Cluster distances and peculiar velocities. Column 3 indicates the source of 
photometric and spectroscopic data, in that order. For each cluster,
$N_{\rm gal}$ indicates the number of galaxies used in the fit. Distances
and peculiar velocities (in the CMB frame) are given in km\,s$^{-1}$. 
The velocity error is a $1\sigma$ 
random error including a contribution from uncertainty in the cluster redshift.} \label{pecvels}
\begin{tabular}{llccrrc}
\\
\hline
& & 
\multicolumn{1}{c}{Data} &
&
&
\multicolumn{1}{c}{\ \ Peculiar \ \ } &
\multicolumn{1}{c}{Velocity} \\
& \multicolumn{1}{l}{Cluster} & 
\multicolumn{1}{c}{Source} &
\multicolumn{1}{c}{\ \ $N_{\rm gal}$ \ \ } &
\multicolumn{1}{c}{\ \ Distance \ \ } &
\multicolumn{1}{c}{\ \ Velocity \ \ } &
\multicolumn{1}{c}{error} \\
\hline
PP ridge 	& 7S21		& 1,1   & 7	& \ \  5448 \ \ \ \ \  &     69 \ \ \ \ \  & 450 \\
		& Pisces  	& 1,1   & 25	& \ \  4583  \ \ \ \ \ &    131 \ \ \ \ \  & 208 \\
		& HMS0122	& 1,1   & 9 	& \ \  4680  \ \ \ \ \ &   --44 \ \ \ \ \  & 352 \\
		& A0262		& 1,1   & 10	& \ \  4787  \ \ \ \ \ &  --259 \ \ \ \ \  & 340 \\
		& A0347		& 1,1   & 8 	& \ \  5590  \ \ \ \ \ &  --277 \ \ \ \ \  & 430 \\
		& Perseus	& 1,1   & 31	& \ \  5176 \ \ \ \ \  &  --136 \ \ \ \ \  & 307 \\
\hline
PP background \ \ \  & J8		& 1,1   & 13	& \ \ 10449  \ \ \ \ \ & --1032 \ \ \ \ \  & 602 \\
		& A2634		& 2,2   & 35	& \ \ 10111 \ \ \ \ \  &  --960 \ \ \ \ \  & 364 \\
\hline
Calibrators 	& A2199		& 2,2   & 36	& \ \  9285  \ \ \ \ \ &  --342 \ \ \ \ \  & 325 \\
		& Coma		& 2,2   & 71	& \ \  7200  \ \ \ \ \ &      0 \ \ \ \ \  & 204 \\
		& A0194		& 3,4+5 & 19	& \ \  4379 \ \ \ \ \  &    743 \ \ \ \ \  & 230 \\
		& A0539		& 3,4   & 22	& \ \  8380  \ \ \ \ \ &    234 \ \ \ \ \  & 402 \\
		& A3381		& 3,4   & 14	& \ \  10883 \ \ \ \ \  &    578 \ \ \ \ \  & 593 \\
		& A3574 	& 3,4   & 7	& \ \  4217 \ \ \ \ \  &    655 \ \ \ \ \  & 369 \\
		& DC2345-28	& 3,5   & 27	& \ \  8645 \ \ \ \ \  &  --174 \ \ \ \ \  & 362 \\
		& Hydra		& 3,4+5 & 18	& \ \  3946 \ \ \ \ \  &     30 \ \ \ \ \  & 239 \\
\hline
\end{tabular}
~\\~\\~\\References for data sources:
1 -- \cite{pp-i}; 2 -- \cite{lu:gu}; 3 -- \cite{jfk1}; 
4 -- \cite{jfk2}; 5 -- \cite{lc88}.
\end{table*}

\begin{table*}
\caption{Comparisons of present results with peculiar velocities measured from the TF
relation by Han \& Mould (1992), and from the FP by Baggley (1996) and Scodeggio et al. (1997).
Baggley's results are based on a subset of the EFAR data. 
Note that our `Pisces' cluster is the `N507 group' of Scodeggio et al.} \label{velcomps}
\begin{tabular}{lcccc}
\\
\hline
\multicolumn{1}{l}{Cluster} & 
\multicolumn{1}{c}{\  This study \  } & 
\multicolumn{1}{c}{\  Han \& Mould (TF) \ } & 
\multicolumn{1}{c}{\  Baggley (FP) \ } &
\multicolumn{1}{c}{\  Scoddegio (FP) \ } \\
\hline
Pisces  	&    131$\pm$208	&    76$\pm$296	&			&   420$\pm$238	\\
HMS0122		&   --44$\pm$352	&   311$\pm$374	&			& 		\\
A0262		&  --259$\pm$340	& --576$\pm$391	&    782$\pm$429		& --157$\pm$203	\\
A2634 		&  --960$\pm$364	& --906$\pm$639	& --978$\pm$967		&  --308$\pm$336	\\
A2199		&  --342$\pm$325	&		&   697$\pm$497		&		\\
Coma		&     (0$\pm$204)	&    80$\pm$428	&    (0$\pm$356)	&    42$\pm$196	\\
A0539		&    234$\pm$402	&    11$\pm$621	&			&		\\
DC2345-28	&  --174$\pm$362	&		&    591$\pm$403		&		\\
Hydra		&     30$\pm$239	&   184$\pm$296	&			&		\\
\hline
\end{tabular}
\end{table*}

\subsection{Bulk flow fits}

As a simple description of the local velocity field, we consider a bulk 
flow model with three free parameters. Fitting for all 16 clusters,
we find a bulk velocity of $430\pm 198$\,km\,s$^{-1}$, in the CMB frame, 
directed towards $l = 264.6^{\circ}, b = -25.6^{\circ}$. Uncertainties in the matching
of spectroscopic systems raise the total error on the bulk-flow amplitude
to 280\,km\,s$^{-1}$.
The bulk motion of this sample of clusters is therefore not statistically
significant.

Fitting a bulk-flow model for only the six PP ridge clusters, no
streaming motion of the supercluster is detected. The fit yields 
a PP bulk-flow of 59$\pm$221\,km\,s$^{-1}$ (toward the Local Group).
The quoted error, includes system matching uncertainties and a contribution
from the uncertainty in setting a velocity zero-point for the sample.

\subsection{Comparison with the predicted velocity field}

A more realistic model of the flow field can be obtained by predicting
the peculiar velocity of each cluster from a suitable redshift survey, 
with $\beta = \Omega^{0.6} / b$ as a single free parameter. Here
we employ the IRAS 1.2 Jy survey, within 12000\,km\,s$^{-1}$,
the IRAS density field being kindly provided to us by M. Strauss.

The best fit to the observed cluster velocities is obtained for
$\beta = 0.94\pm 0.48$. While this represents a marginal detection of 
$\beta$, the errors are too large for useful constraints to be derived
from the present sample.

After subtraction of the IRAS-predicted peculiar velocities
from the observed motions, the residual (CMB frame) bulk motion of
the sample is
$383\pm 294$\,km\,s$^{-1}$, directed towards $l = 313.2^{\circ}, 
b = -26.4^{\circ}$.
We conclude from this non-detection, that mass concentrations beyond
12000\,km\,s$^{-1}$ are not required to explain the observed velocities.

\section{Conclusions} \label{conclusions}

We have completed a new survey of cluster motions 
in the PP region, using the FP as a distance indicator. Careful
attention has been paid to the construction of a homogeneous, merged
dataset, and to
accounting for systematic uncertainties in this procedure. 

We derive an insignificant bulk 
motion (--60$\pm$220 km\,s$^{-1}$ towards the Local
Group) for 6 
clusters in the main ridge of the supercluster.  
This result is in marginal conflict with Tully--Fisher surveys
by \cite{w91}, \cite{ha:mo} and \cite{cf93}. Two clusters in the background of PP 
show evidence for `backside infall' into the supercluster.

Comparing the observed cluster motions with the velocity field predicted
from the IRAS 1.2Jy survey, we find a best fit for $\beta = 1.0\pm 0.5$. 
There is no evidence for residual bulk flows generated by sources beyond
the IRAS density field limit.

A new, all-sky, survey of $\sim$\,50 Abell clusters within 12000\,km\,s$^{-1}$ is
currently in progress (see Smith et al. 1997b, this volume).  
This improved sample, will yield
useful constraints on $\beta$, in addition to a reliable measurent of 
the bulk motion.

\end{document}